\documentclass[aps,prc,preprint,groupedaddress]{revtex4-1}
\usepackage{graphicx}
\usepackage{epsfig}
\usepackage{geometry}
\usepackage{fancyhdr,amsmath,amssymb,hyperref}
\usepackage{longtable}
\begin{document}

%\preprint{}

\title{Multi-nucleon transfer in the interaction of 977 MeV and 1143 MeV $^{204}$Hg with $^{208}$Pb}


\author{V. V.  Desai, A. Pica,  W. Loveland, J.S. Barrett}
%\thanks{}
\affiliation{Department of Chemistry, Oregon State University, Corvallis, Oregon 97331 USA}
\author{E.A. McCutchan}
\affiliation{National Nuclear Data Center, Brookhaven National Laboratory, Upton, New York 11973, USA}
\author{S. Zhu} 
\altaffiliation{Present address National Nuclear Data Center, Brookhaven National Laboratory, Upton, New York 11973, USA} 
\author{A. D. Ayangeakaa}
\altaffiliation{Present address:  United States  Naval Academy, Annapolis,  Maryland 21402 USA}
\author{M. P.  Carpenter, J.P. Greene,  T. Lauritsen}
\affiliation{Physics Division, Argonne National Laboratory, Argonne, Illinois 60439 USA}
\author {R.V.F. Janssens}
\affiliation{Department of Physics and Astronomy, University of North Carolina at Chapel Hill, Chapel Hill, North Carolina 27599 USA and Triangle Universities Nuclear Laboratory, Duke University, Durham, North Carolina 27708 USA}
\author{B. M. S.  Amro}
\affiliation{Dept. of Physics, University of  Massachusetts  Lowell, Lowell MA 01854 USA}
\author{W. B. Walters}
\affiliation{Dept. of Chemistry, University of Maryland, College Park, MD}


\date{\today}

\begin{abstract}
A previous study of symmetric collisions of massive nuclei has  shown that current models of multi-nucleon transfer (MNT) reactions do not adequately describe the transfer product yields.  To gain further insight into this problem, we have measured the yields of MNT products  in the interaction of 977 (E/A = 4.79 MeV) and 1143 MeV (E/A = 5.60 MeV) $^{204}$Hg with $^{208}$Pb.  We find that the yield of multi-nucleon transfer products are similar in these two reactions and are substantially lower than those observed in the reaction of 1257 MeV (E/A = 6.16 MeV) $^{204}$Hg + $^{198}$Pt.  We compare our measurements with the predictions of the GRAZING-F, di-nuclear systems (DNS) and improved quantum molecular dynamics (ImQMD) models.  For the observed isotopes of the elements Au, Hg, Tl, Pb and Bi, the measured values of the MNT cross sections are orders of magnitude larger than the predicted values.  Furthermore, the various models predict the formation of nuclides near the N=126 shell, which are not observed.
\end{abstract}


\maketitle

\section{Introduction}

  Multi-nucleon transfer (MNT) reactions are thought to be useful paths for synthesizing new n-rich heavy nuclei \cite{zg2,zg3} and as possible paths for synthesizing nuclei near the $\mathit{N}$=126 shell closure (of interest to the studies of r-process nucleosynthesis \cite{zg}).   Some of the most interesting of these reactions involve the near symmetric collisions of massive nuclei, such as $^{238}$U + $^{248}$Cm.  In this regard, the recent result of Welsh et al. \cite{tim} is somewhat disturbing.  Welsh et al. measured the yields of several nuclides  from the near symmetric reaction of 1257 MeV $^{204}$Hg  with $^{198}$Pt.  They found that the yields of the transfer products were significantly larger, even for small transfers, than  those predicted by typical models for MNT reactions, such as GRAZING, the DNS model and the ImQMD model.  While it  is encouraging to see the larger than expected yields of the MNT products, it is vexing that we are unable to predict these yields even for the smallest transfers,  let alone the larger ones. Accordingly we undertook an investigation, reported in this paper, of the projectile-like fragments (PLFs) and target-like fragments (TLFs) yields in another symmetric reaction, the reaction of 977 and 1143 MeV $^{204}$Hg with $^{208}$Pb. By making this investigation, we hope to check whether there are some special features of near symmetric collisions that affect the MNT yields.
 
\section{Experimental}
The experimental method used was similar to that of Desai et al. \cite{des}. Using the Gammasphere facility of the Argonne National Laboratory, beams of $^{204}$Hg struck targets of $^{208}$Pb.  For the irradiation at 977 MeV, the effective target thickness was 19.5 mg/cm$^{2}$ and the total bombardment time was 1404 min.  For the irradiation at 1143 MeV, the effective target thickness was 28.0 mg/cm$^{2}$ and the total bombardment time was 2632 min.  (In the 977 MeV study the actual beam energy was 1360 MeV.  The incident beam loses energy as it goes through the target and after traversing 19.5 mg/cm$^{2}$ of $^{208}$Pb, the beam energy drops below the interaction barrier of 586.7 MeV.  Thus the `effective' target thickness was 19.5 mg/cm$^{2}$ while the physical target thickness was 48 mg/cm$^{2}$) .  In the higher energy irradiation (1143 MeV), the incident beam energy was 1700 MeV, the physical target thickness was 44 mg/cm$^{2}$, and the effective target thickness was 27.7 mg/cm$^{2}$.  The intensity of the beam was monitored periodically by inserting  a suppressed Faraday cup in the beam line in front of the target. The beam intensities were 3.07 x 10$^{10}$ and 3.17 x 10$^{10}$ ions/min for the lower and higher energy irradiations, respectively.  The lower energy irradiation was performed in May 2015,  while the higher energy irradiation was performed in April, 2016.  

At the end of each irradiation,  the target was removed from Gammasphere and $\gamma$-ray spectroscopy of the target radioactivities was carried out using a well-calibrated Ge detector in The Center for Accelerator Target Science (CATS) Counting Laboratory.  The total observation period for the lower energy was 5 days, during which 25 measurements of target radioactivity were made.  The total observation time for the higher energy was 7 days during which 23 measurements of the sample were made.  The analysis of these Ge $\gamma$-ray decay spectra was carried out using the FitzPeaks \cite{jim} software.  The end of bombardment (EOB) activities of the nuclides were used to calculate absolute production cross sections, taking into account the variable beam intensities using standard equations for the growth and decay of radionuclides during irradiation \cite {FKMM}.  These measured absolute nuclidic production cross sections are tabulated in Tables 1 and 2.  These cross sections represent ``cumulative" yields;, i.e., they have not been corrected for the effects of precursor beta decay. These cumulative yields are the primary measured quantity in this experiment.  

To correct for precursor beta decay, we have assumed that the beta-decay corrected independent yield cross sections for a given species, $\sigma$(Z,A), can be represented as a histogram that lies along a Gaussian curve
\begin{equation}
\sigma (Z,A)=\sigma (A)\left[ 2\pi C_{Z}^{2}(A)\right] ^{-1/2}\exp\left[ \frac{-(Z-Z_{mp})^{2}}{2C_{Z}^{2}(A)}\right] 
\end{equation}
where $\sigma$(A) is the total isobaric yield (the mass yield), C$_{Z}$(A) is the Gaussian width parameter for mass number A, and Z$_{mp}$(A) is the most probable atomic number for that A.  Given this assumption, the beta-decay feeding correction factors for cumulative yield isobars can be calculated,  once the centroid and width of the Gaussian function are known.  

To uniquely specify $\sigma$(A), C$_{Z}$(A), and Z$_{mp}$(A), one would need to measure three independent yield cross sections for each isobar.  This is difficult and generally not feasible for most isobars.  Instead, one assumes that  the value of  $\sigma$(A) varies smoothly and slowly as a function of mass number and is roughly constant within any A range when determining C$_{Z}$(A) and Z$_{mp}$(A). The measured nuclidic formation cross sections are then placed in  groups according to mass number.    We assume that the charge distributions of neighboring isobaric chains are similar and radionuclide yields from a limited mass region can be used to determine a single charge distribution curve for that mass region.  One can then use the laws of radioactive decay to iteratively correct the measured cumulative formation cross sections for precursor decay.  These ``independent yield" cross sections are also tabulated in Tables 1 and 2.  The cumulative and independent yield cross sections are similar due to the fact that, without an external separation of the reaction products by Z or A, one most likely detects only a single or a few nuclides for a given isobaric chain,  and these nuclides are located near the maximum of the Gaussian yield distribution.  The uncertainties in the calculated ``independent yield" cross sections deduced in this manner have been examined by Morrissey et al. \cite{djm} and they have found a systematic uncertainty of ~$\pm$ 30 \% associated with this procedure.

\section{Results and Discussion}

Due to the nearly symmetric character of the $^{204}$Hg + $^{208}$Pb reaction, separation of the products into projectile-like fragments (PLFs) and target-like fragments (TLFs) is not meaningful.  While  some models for these reactions classify fragments as PLFs and TLFs, we have summed all of the predicted yields to give ``fragment yields".  

\subsection{Comparison with phenomenological models}

A well-known model for predicting the cross sections for transfer products is GRAZING, a semi-classical model due to Pollarolo and Winther \cite{nanni, winther}. GRAZING uses a semi-classical model of the reacting ions moving on classical trajectories with quantum calculations of the probability of excitation of collective states and of nucleon transfer. This model describes few nucleon transfers \cite{corradi} well. It has been employed to describe the production of projectile like fragments (PLFs) involving transfers of 4�5 nucleons in the asymmetric reaction of $^{136}$Xe with $^{238}$U, where the predictions of this model agree well with measurements \cite{vogt}.  Yanez and Loveland \cite{yanez} have published a variant of the GRAZING code, called GRAZING-F, which takes into account the decay of the MNT primary fragments by both fission and neutron emission.The measured and predicted (GRAZING-F) values for selected nuclides  are shown in Figures 1 -10.

Another model for predicting the yields of MNT products is the dinuclear system  (DNS) model which is described in \cite {LZ1,Wen}.  Unlike the GRAZING model, this model focuses on the more central collisions, in which there is considerable overlap between the colliding nuclei.  The GRAZING and DNS models are, thus, complementary.  The predictions of the DNS model are compared to the measured data  in Figures 1-10, as well.

A third model for predicting MNT yields that has been quite successful \cite{tim,des} is the Improved Quantum Molecular Dynamics (ImQMD) model \cite{li1,li2}.  This model has been shown to describe MNT yields in a wide variety of reactions.  The predictions of this model are also compared to the experimental data in Figures 1-10

One's first impression from Figures 1 -10 is that the increase in beam energy from 977 to 1143 MeV has a small effect on the measured cross sections.

\begin{figure}
\includegraphics[scale=0.7]{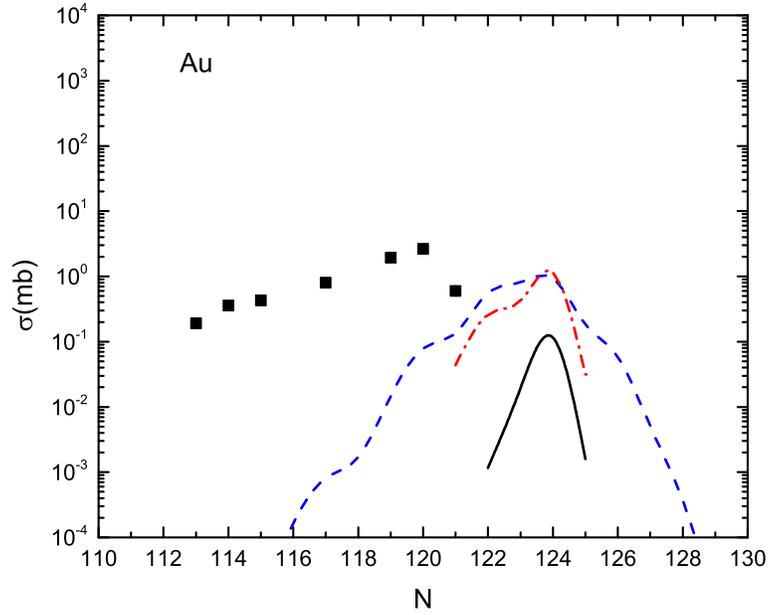}
 \caption{A comparison of the  predicted (GRAZING-F, DNS and ImQMD) yields and the measured yields of the Au isotopes formed in the reaction of 977 MeV $^{204}$Hg with $^{208}$Pb.  The solid squares represent the experimental data, while the solid , dashed  and the dash-dot lines represent the predictions of the GRAZING-F, DNS and ImQMD models, respectively.}
 \end{figure}

 \begin{figure}
\includegraphics[scale=0.7]{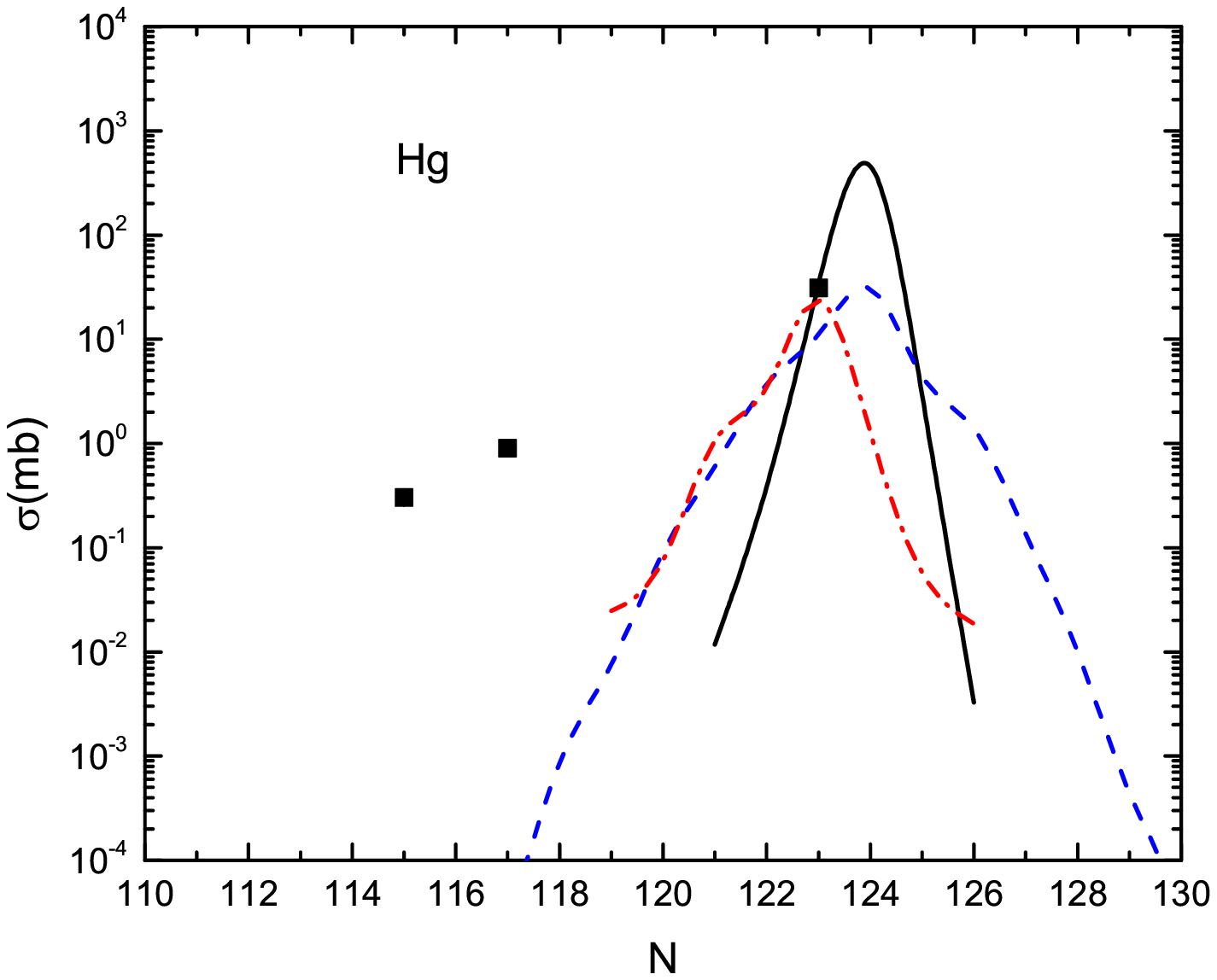}
\caption{A comparison of the  predicted and measured yields of the Hg isotopes formed in the reaction of 977 MeV $^{204}$Hg with $^{208}$Pb.  See Figure 1 for the meaning of the symbols.}
 \end{figure}

 \begin{figure}
\includegraphics[scale=0.7]{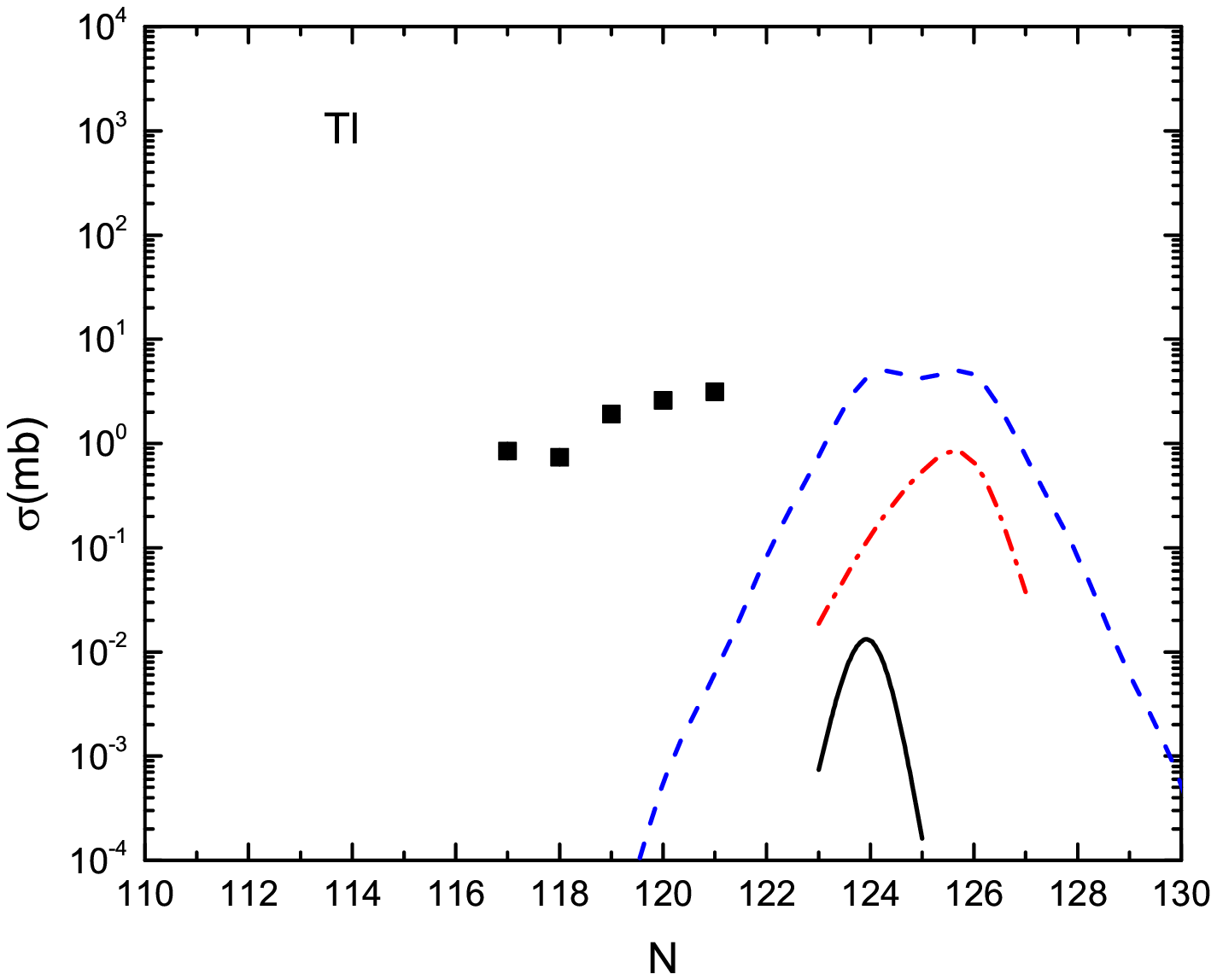}
\caption{A comparison of the  predicted and measured yields of the Tl isotopes formed in the reaction of 977 MeV $^{204}$Hg with $^{208}$Pb.  See Figure 1 for the meaning of the symbols.}
\end{figure}

 \begin{figure}
\includegraphics[scale=0.7]{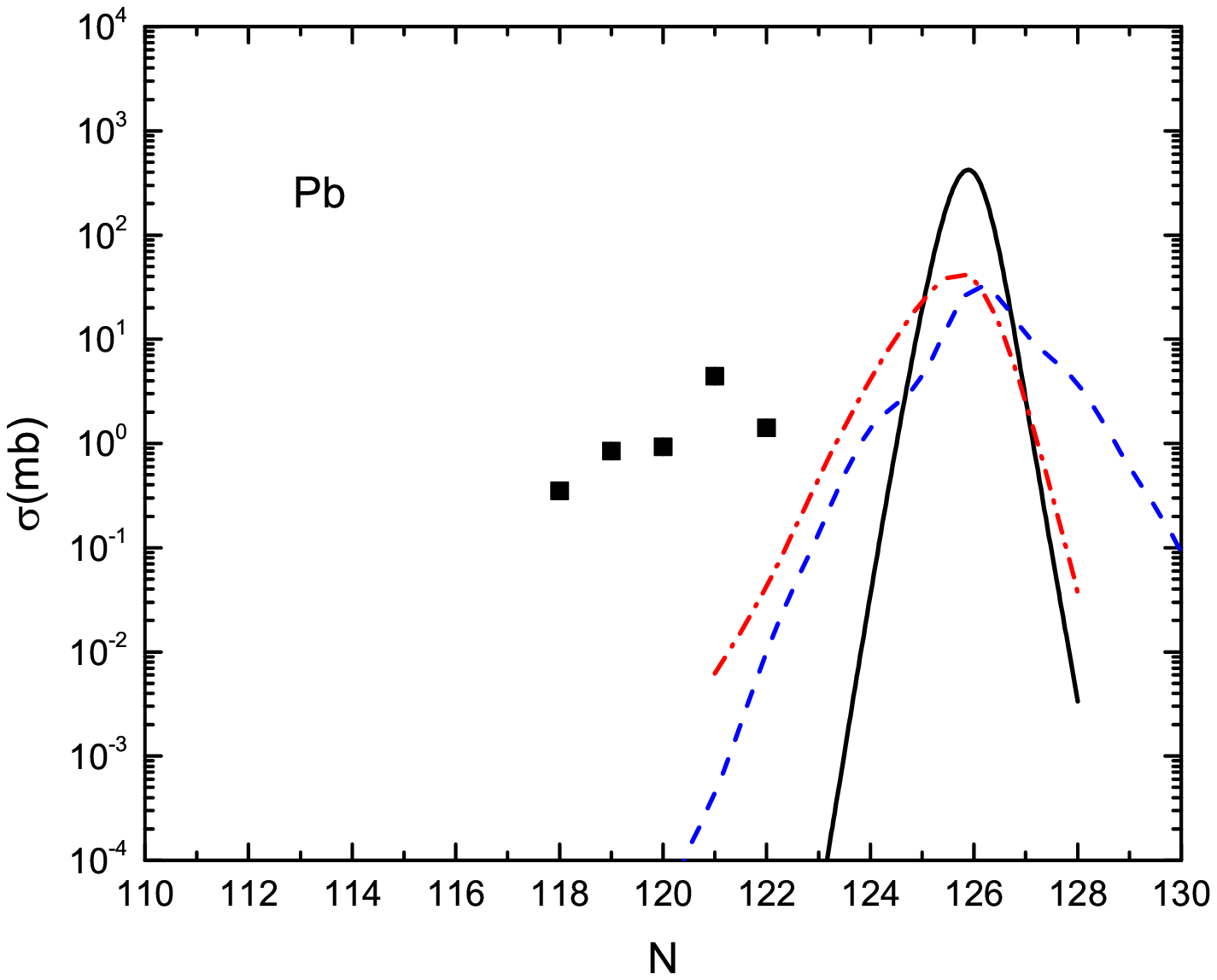}
\caption{A comparison of the  predicted and measured yields of the Pb isotopes formed in the reaction of 977 MeV $^{204}$Hg with $^{208}$Pb.  See Figure 1 for the meaning of the symbols.}
\end{figure}

\begin{figure}
\includegraphics[scale=0.7]{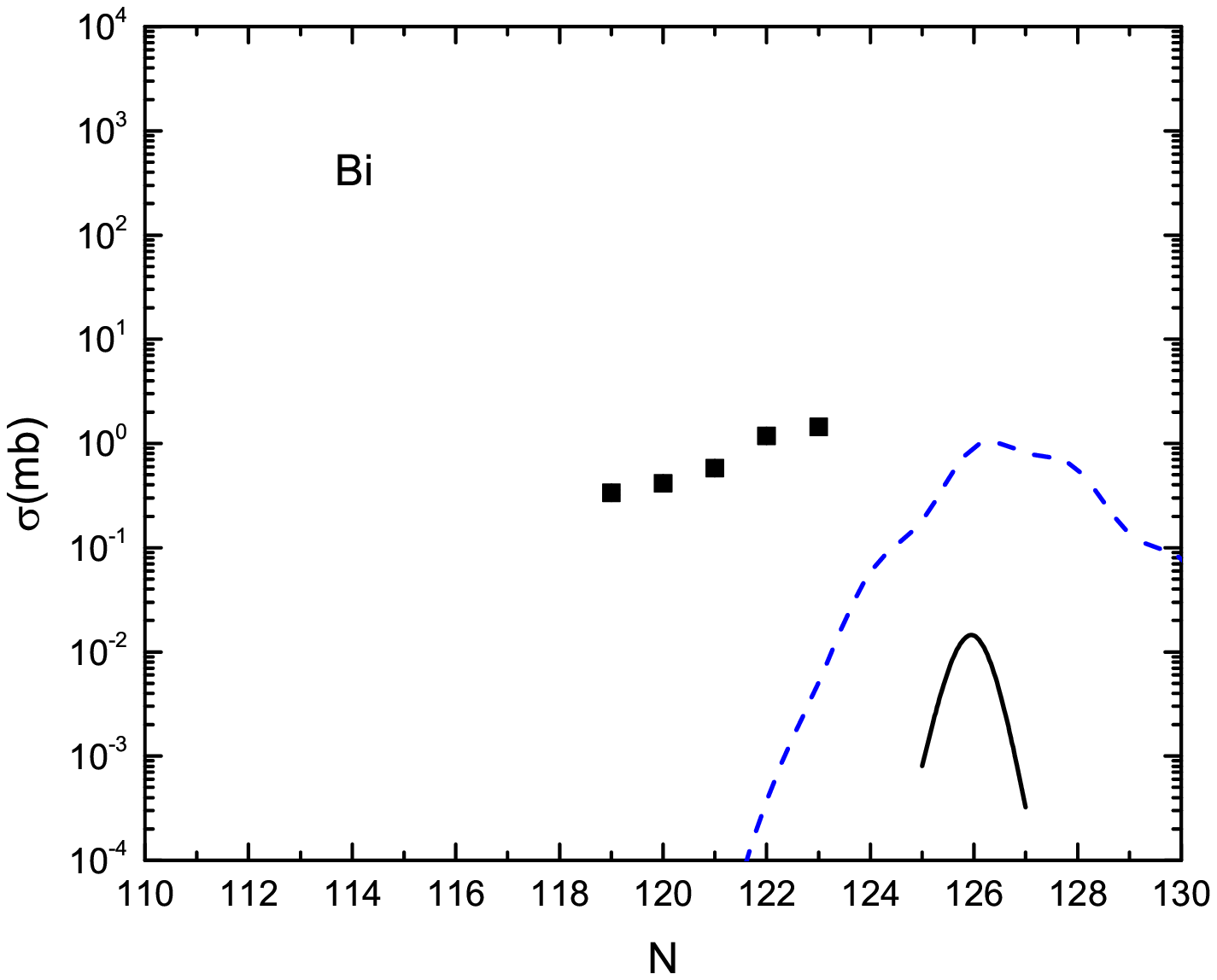}
\caption{A comparison of the  predicted and measured yields of the Bi isotopes formed in the reaction of 977 MeV $^{204}$Hg with $^{208}$Pb.  See Figure 1 for the meaning of the symbols.}
\end{figure}

\begin{figure}
\includegraphics[scale=0.7]{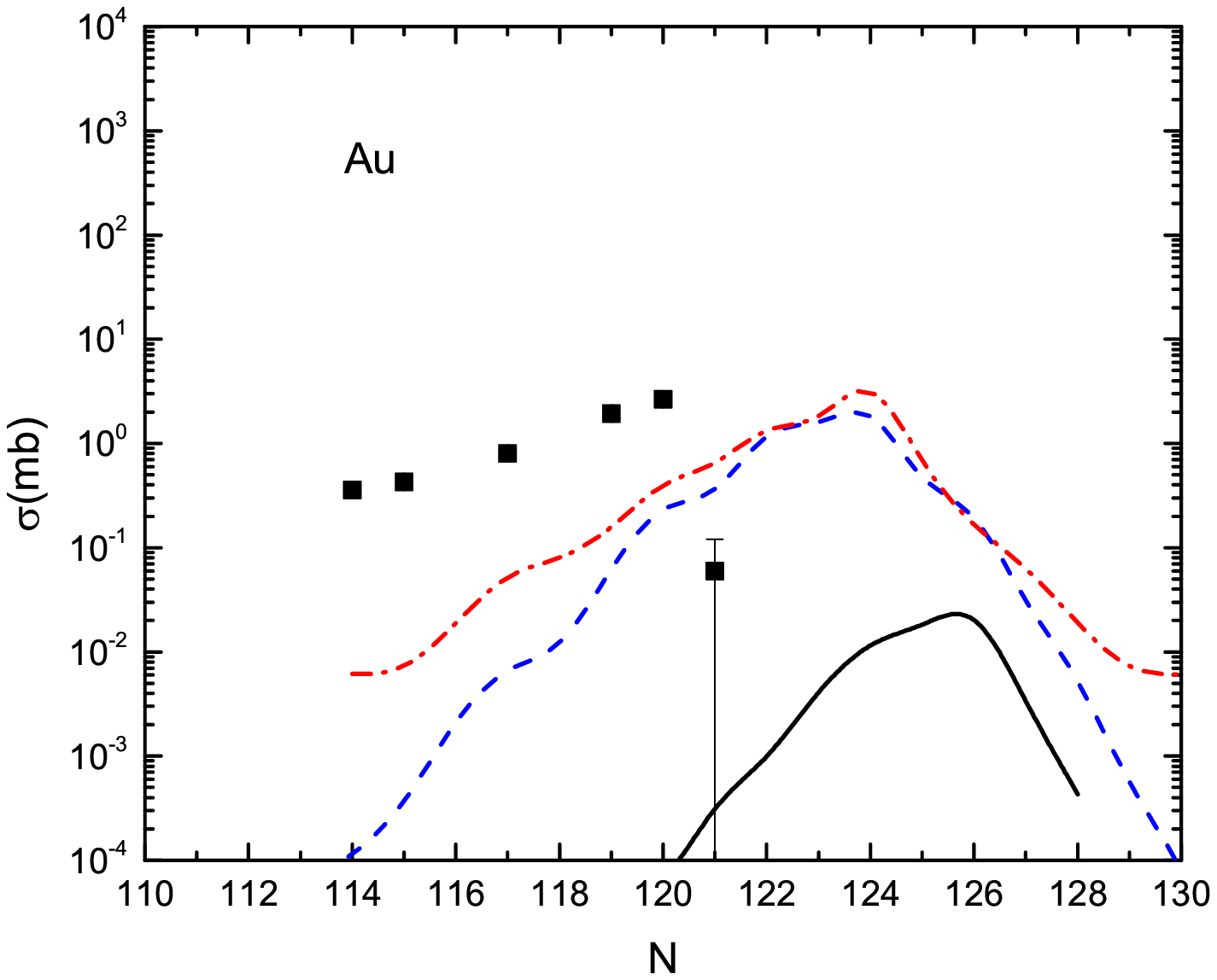}
 \caption{A comparison of the  predicted (GRAZING-F, DNS and ImQMD) yields and the measured yields of the Au isotopes formed in the reaction of 1143 MeV $^{204}$Hg with $^{208}$Pb.  The solid squares represent the experimental data, while the solid ,  dashed  and the dash-dot lines represent the predictions of the GRAZING-F, DNS and ImQMD models, respectively.}
 \end{figure}

 \begin{figure}
\includegraphics[scale=0.7]{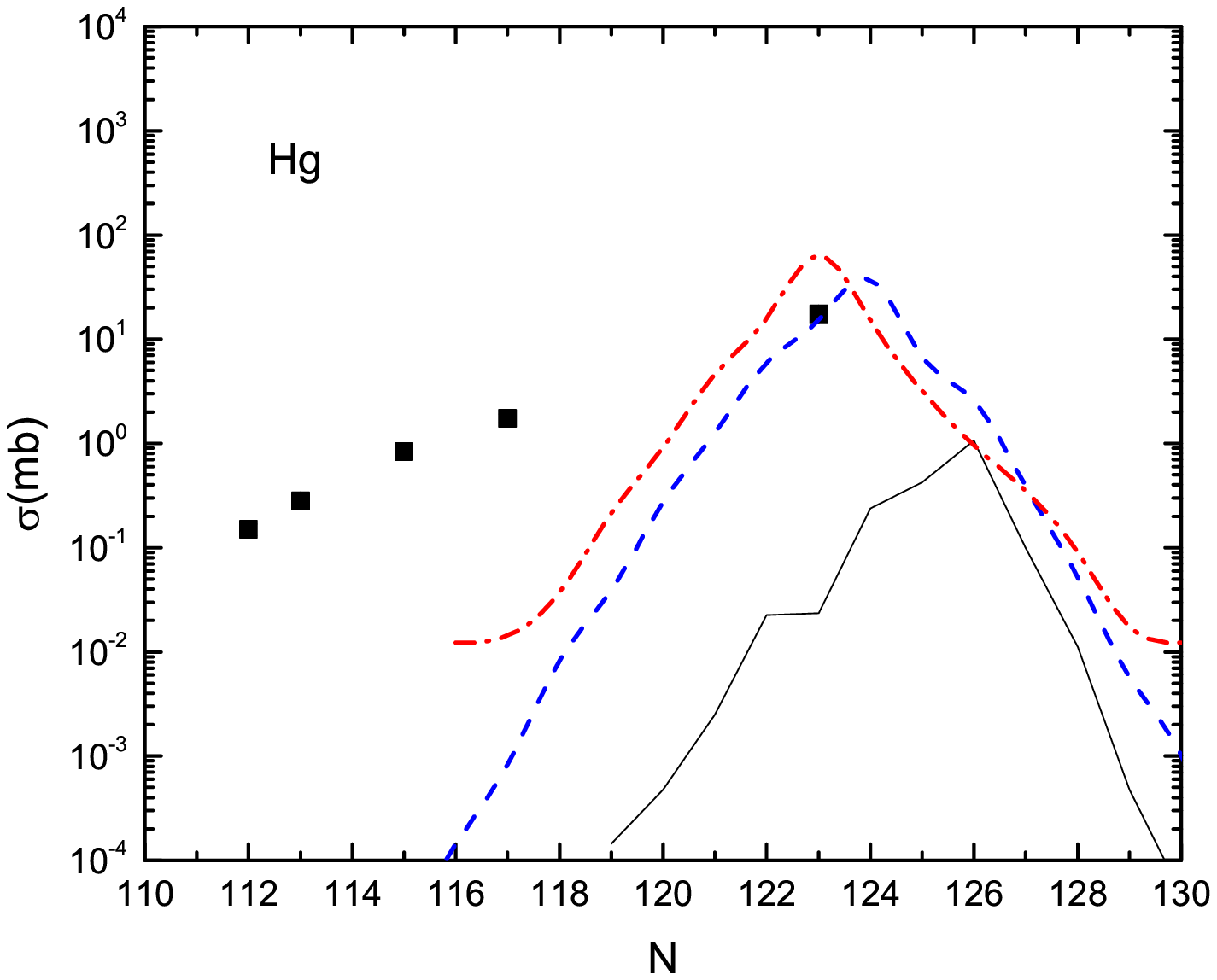}
\caption{A comparison of the  predicted and measured yields of the Hg isotopes formed in the reaction of 1143 MeV $^{204}$Hg with $^{208}$Pb.  See Figure 6 for the meaning of the symbols.}
 \end{figure}

 \begin{figure}
\includegraphics[scale=0.7]{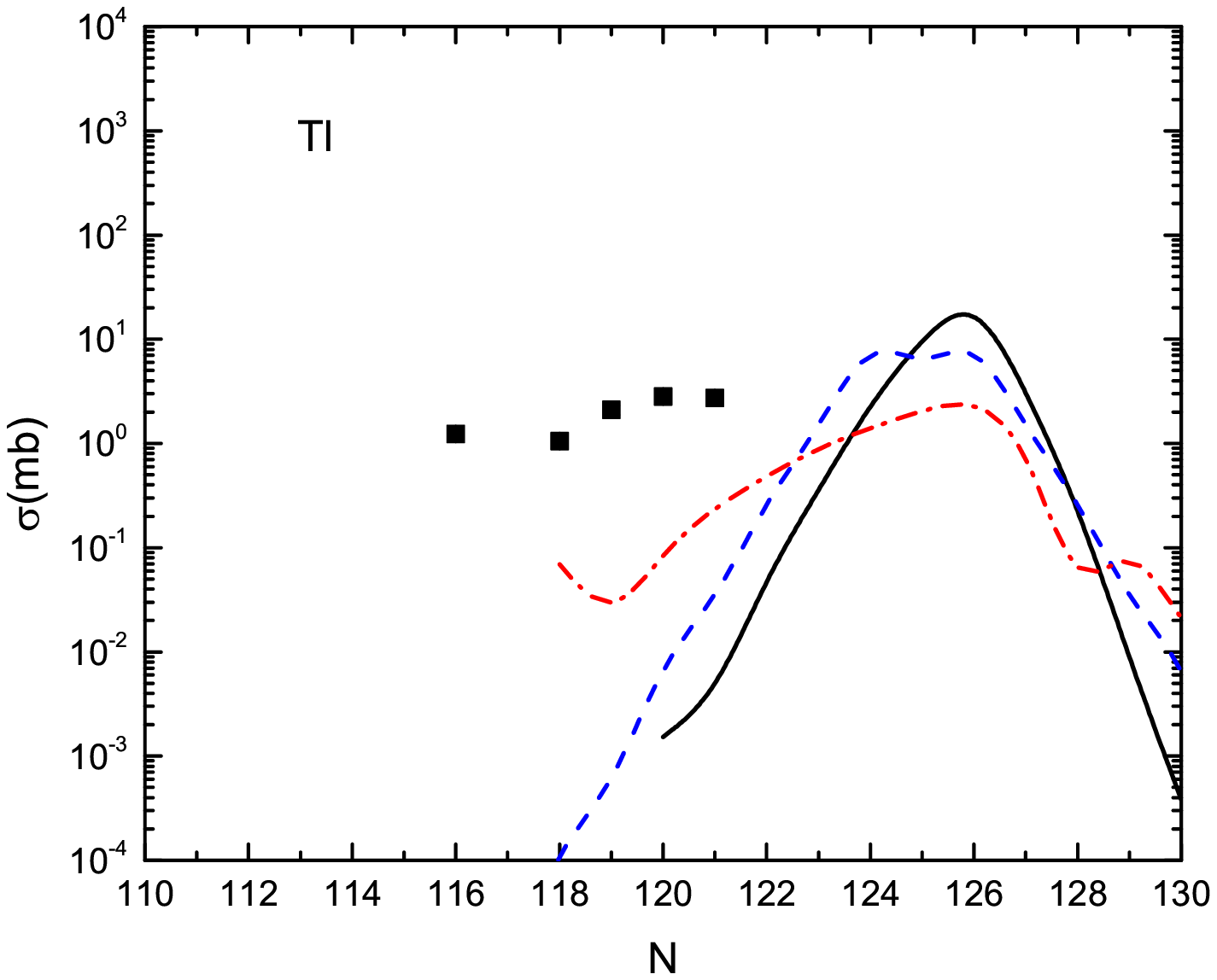}
\caption{A comparison of the  predicted and measured yields of the Tl isotopes formed in the reaction of 1143 MeV $^{204}$Hg with $^{208}$Pb.  See Figure 6 for the meaning of the symbols.}
\end{figure}

 \begin{figure}
\includegraphics[scale=0.7]{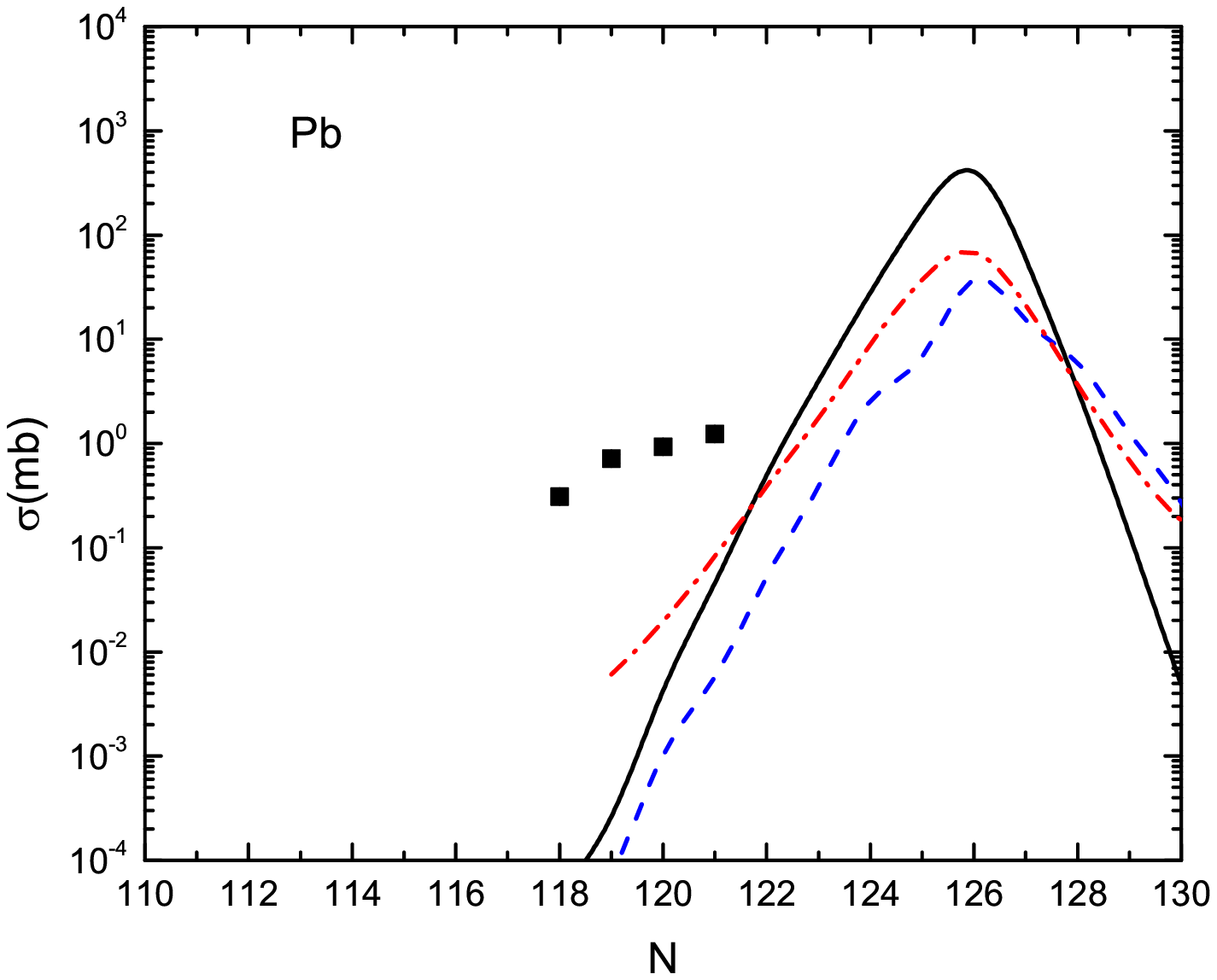}
\caption{A comparison of the  predicted and measured yields of the Pb isotopes formed in the reaction of 1143 MeV $^{204}$Hg with $^{208}$Pb.  See Figure 6 for the meaning of the symbols.}
\end{figure}

\begin{figure}
\includegraphics[scale=0.7]{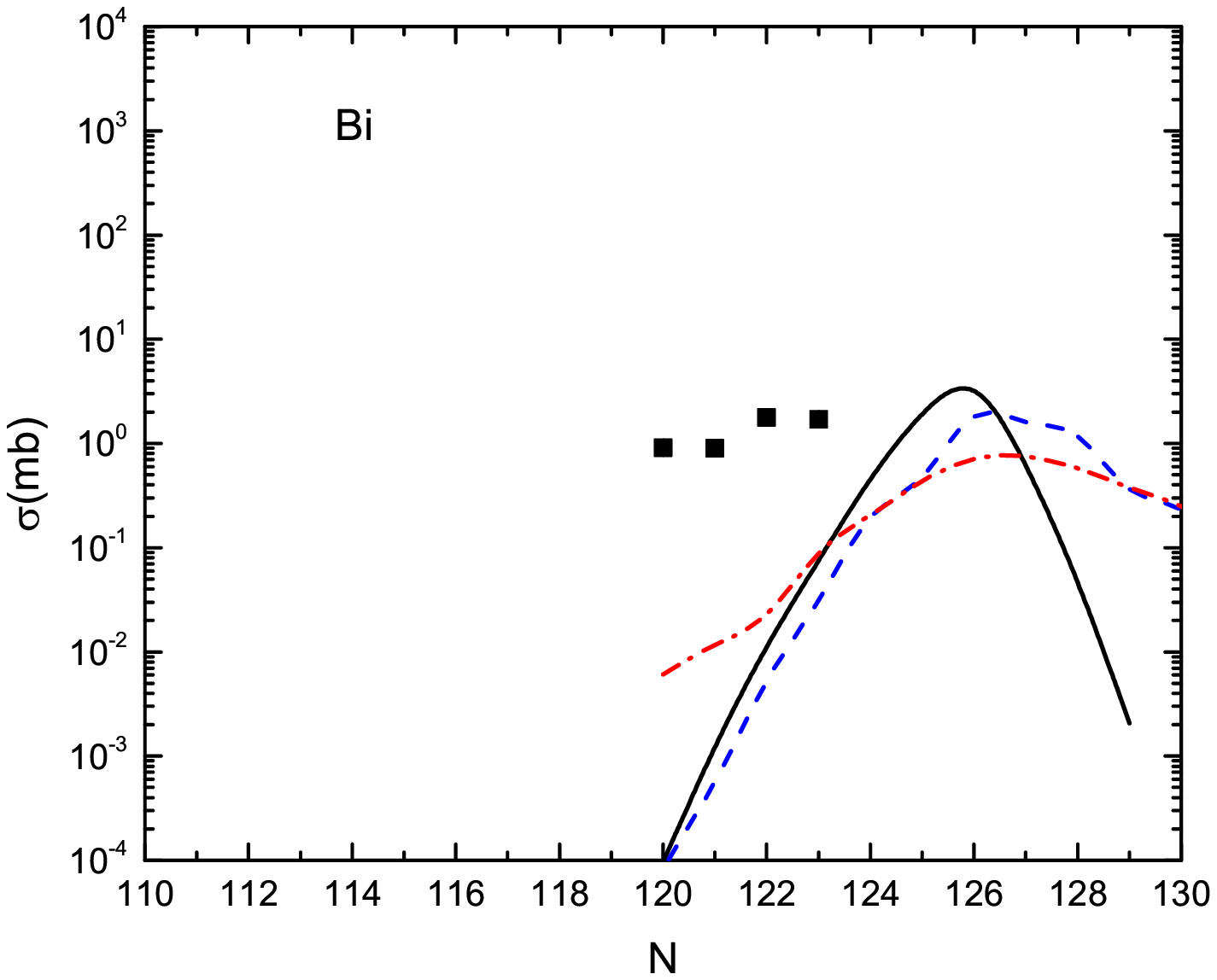}
\caption{A comparison of the  predicted and measured yields of the Bi isotopes formed in the reaction of 1143 MeV $^{204}$Hg with $^{208}$Pb.  See Figure 6 for the meaning of the symbols.}
\end{figure}

In Figures 1-10, the observed MNT products are more neutron-deficient than those predicted by the models, with the exception of $^{203}$Hg.  The observed yields are orders of magnitude larger than those predicted by the various models.  As the atomic number of the elements increases, the various models predict large yields for the nuclei near the N=126 shell-- a prediction that is not consistent with the measurements.  One can be encouraged or discouraged by this situation.The fact that the measured cross sections are larger than the predicted cross sections is encouraging for using MNT reactions to synthesize new n-rich nuclei, but the inability to see nuclei near the N=126 shell might indicate that these symmetric reactions are not a suitable path to these very n-rich nuclei.   For all the models there is an interesting "odd-even" effect with the atomic numbers of the MNT products.  The even Z nuclides ( Hg and Pb) show higher yields than the odd Z nuclides (Au, Tl, Bi) but one must remember Hg and Pb were the projectile and target, respectively.  

If we compare the measured cross sections from this work (977 and 1143 MeV $^{204}$Hg  + $^{208}$Pb) with the measurements of Welsh et al. \cite{tim} for (1257 MeV $^{204}$Hg + $^{198}$Pt)(Figure 11)  we observe similar yield patterns except that the cross sections for the higher energy reaction (1257 MeV $^{204}$Hg + $^{198}$Pt) are substantially greater.

\begin{figure}
\includegraphics{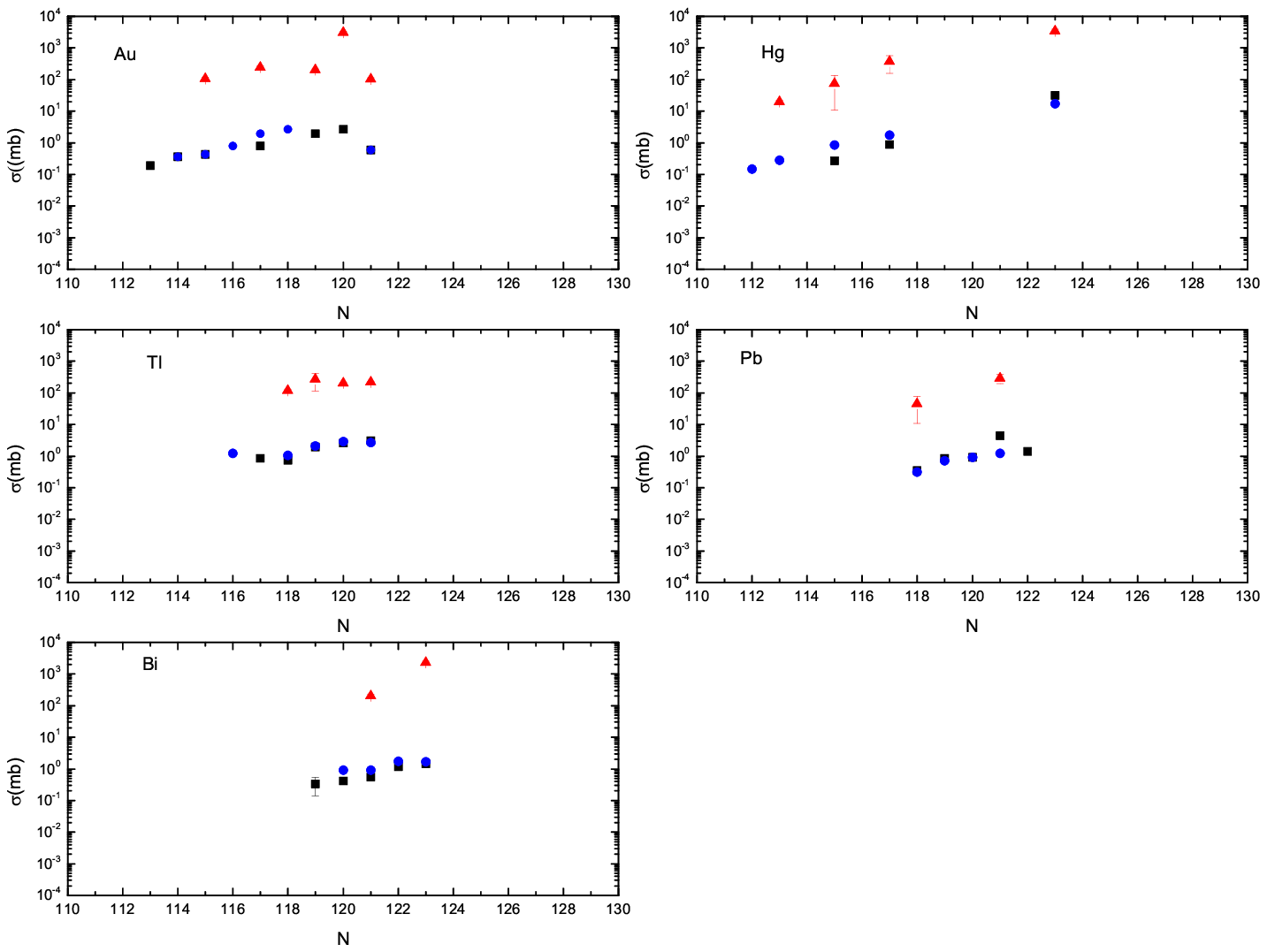}
\caption{A comparison of the yields of transfer products in the reaction of 997 MeV $^{203}$Hg  + $^{208}$Pb (this work, black squares)), 1143 MeV $^{203}$Hg  + $^{208}$Pb (this work, blue circles)) and the reaction of 1257 MeV $^{203}$Hg  + $^{198}$Pt \cite{tim} (red triangles) }
\end{figure}

\section{Conclusions}

 What have we learned from this experiment?  We found that:  (a) There is very little change in the yields of the MNT transfer products when the beam energy is raised from 977 MeV to 1143 MeV. (E/A = 4.79 to 5.60 MeV/A)(b) Comparing our results to those of Welsh et al. \cite{tim}, we find that raising the projectile from 977 or 1143 MeV to 1267 MeV increases the MNT production cross sections by about two orders of magnitude. (c) The frequently used models for MNT collisions (GRAZING-F, DNS and ImQMD) fail to describe these symmetric collisions, a situation similar to that observed by Welsh et al. \cite {tim}.  This is not a trivial observation as symmetric reactions like U + Cm, etc are frequently cited as pathways to n-rich heavy nuclei.

\section{Acknowledgements}
This material is based upon work supported in part  by the U.S. Department of Energy, Office of Science, Office of Nuclear Physics under award numbers DE-FG06-97ER41026 (OSU), DE-FG02-97ER41041 (UNC), DE-FG02-97ER41033 (TUNL), DE-FG02-94ER40848 (UMassLowell) and contract numbers DE-AC02-06CH11357 (ANL) and DE-AC02-98CH10886(BNL).  This research used resources of ANL's ATLAS facility, which is a DOE Office of Science User facility.

\begin{longtable}{|c|c|c|}
\caption{Fragment cumulative and independent yields for  the reaction of  $^{204}$Hg + $^{208}$Pb at E$_{lab}$ = 977 MeV.} \\
\hline
\textbf{Isotope}& \textbf{$\sigma$$_{CY}$ (mb)}& \textbf{$\sigma$$_{IY}$ (mb)} \\
\hline 
\endfirsthead 
\multicolumn{3}{c}{\tablename\ \thetable\ -- \textit{Continued from previous page}} \\
\hline
\textbf{Isotope}& \textbf{$\sigma$$_{CY}$ (mb)}& \textbf{$\sigma$$_{IY}$ (mb)} \\
\hline
\endhead
\hline \multicolumn{3}{c}{\textit{Continued on next page}} \\
\endfoot
\hline
\endlastfoot
$^{72}$Ga&0.078 $\pm$0.009 &0.073  $\pm$0.009 \\
$^{81}$Rb$^{m}$&0.029 $\pm$0.020 & 0.024  $\pm$ 0.016 \\
$^{85}$Kr & 0.052 $\pm$ 0.017 & 0.044  $\pm$ 0.014 \\
\hline
$^{91}$Sr & 0.080 $\pm$ 0.023 & 0.070 $\pm$ 0.020 \\
$^{92}$Sr & 0.183 $\pm$ 0.012 & 0.183 $\pm$ 0.018  \\
\hline
$^{97}$Zr & 0.205 $\pm$ 0.010 & 0.205 $\pm$ 0.021 \\
\hline
$^{90}$Nb & 0.159 $\pm$ 0.029 & 0.143 $\pm$ 0.026 \\
\hline
$^{99}$Mo & 0.178 $\pm$ 0.001  & 0.178 $\pm$ 0.018 \\
\hline
$^{93}$Tc & 0.21 $\pm$ 0.06 & 0.21 $\pm$  0.06 \\
$^{99}$Tc & 0.030 $\pm$ 0.009 & 0.023 $\pm$ 0.007 \\
\hline
$^{97}$Ru & 0.751 $\pm$ 0.011 & 0.691 $\pm$ 0.069 \\
$^{103}$Ru & 1.204 $\pm$ 0.082 & 0.773 $\pm$ 0.077 \\
\hline
$^{112}$Ag &  0.232  $\pm$  0.059 & 0.182 $\pm$  0.046\\
\hline 
$^{115}$Cd & 0.130 $\pm$ 0.004 & 0.097 $\pm$ 0.010 \\
\hline
$^{117}$Sb& 0.585 $\pm$ 0.027 & 0.505 $\pm$ 0.051  \\
\hline
$^{131}$I&0.082 $\pm$ 0.048 & 0.073 $\pm$ 0.043 \\
\hline
$^{125}$Xe&0.187 $\pm$ 0.011 & 0.161 $\pm$ 0.016\\
$^{135}$Xe&0.079 $\pm$ 0.009 & 0.074 $\pm$ 0.008\\
\hline
$^{140}$Ba & 0.195 $\pm$ 0.020 & 0.180 $\pm$ 0.018 \\
\hline
$^{143}$Ce&  0.099 $\pm$  0.009  & 0.084 $\pm$  0.008    \\
\hline
$^{156}$Sm&  0.278 $\pm$  0.035  & 0.278  $\pm$  0.035  \\
\hline
$^{154}$Tb&  0.737 $\pm$  0.025  & 0.341  $\pm$  0.034  \\
\hline
$^{167}$Ho&  0.124 $\pm$  0.019  & 0.106   $\pm$  0.016  \\
\hline
$^{169}$Lu &  0.221 $\pm$ 0.015 & 0.221 $\pm$ 0.022 \\
\hline
$^{180}$Hf$^{m}$& 0.095 $\pm$  0.010 & 0.095 $\pm$  0.010  \\
\hline
$^{181}$Re& 1.39$\pm$ 1.05 & 1.39$\pm$ 1.05  \\
$^{182}$Re& 1.883$\pm$ 0.093  & 1.883 $\pm$  0.188 \\
$^{188}$Re& 0.219$\pm$ 0.037 & 0.219 $\pm$  0.037 \\
\hline
$^{191}$Os& 0.594 $\pm$ 0.031 & 0.281 $\pm$ 0.028  \\
\hline
$^{188}$Ir& 0.192 $\pm$  0.016 & 0.187 $\pm$  0.019   \\
$^{190}$Ir&  0.137 $\pm$  0.005   & 0.068  $\pm$  0.007   \\
\hline
$^{195}$Pt& 0.918 $\pm$ 0.048  & 0.742 $\pm$  0.074  \\

\hline
$^{192}$Au & 0.221$\pm$ 0.007 &  0.190 $\pm$  0.019    \\
$^{193}$Au&  0.430 $\pm$  0.072  &  0.357 $\pm$    0.060  \\
$^{194}$Au&  0.426 $\pm$  0.043  & 0.426 $\pm$    0.043  \\
$^{196}$Au&  1.652   $\pm$  0.028  &  0.804$\pm$   0.080    \\
$^{198}$Au&  2.414 $\pm$ 0.010  & 1.930  $\pm$  0.193  \\
$^{199}$Au&  3.25 $\pm$   0.02 &  2.65 $\pm$    0.26   \\
$^{200}$Au$^{m}$ & 0.598 $\pm$ 0.026 &  0.598 $\pm$ 0.060  \\
\hline
$^{195}$Hg$^{m}$& 0.303 $\pm$  0.011 & 0.273 $\pm$ 0.027   \\
$^{197}$Hg& 0.897 $\pm$  0.140  &  0.897 $\pm$   0.140   \\
$^{203}$Hg & 38.1 $\pm$ 1.6 & 31.2 $\pm$ 3.1 \\
\hline
$^{198}$Tl&  0.963 $\pm$  0.080 & 0.847  $\pm$ 0.085    \\
$^{199}$Tl&  0.870 $\pm$  0.018 & 0.741  $\pm$   0.074  \\
$^{200}$Tl&  2.21 $\pm$  0.14 &  1.91 $\pm$  0.19   \\
$^{201}$Tl&  3.27 $\pm$   0.03  & 2.59 $\pm$  0.26    \\
$^{202}$Tl& 3.14  $\pm$  0.03  & 3.14 $\pm$  0.31    \\
\hline
$^{200}$Pb& 0.388$\pm$ 0.012 & 0.352 $\pm$ 0.035    \\
$^{201}$Pb&  0.962 $\pm$  0.077  &  0.845 $\pm$  0.085   \\
$^{202}$Pb$^{m}$& 0.926 $\pm$  0.015 &  0.926 $\pm$   0.093   \\
$^{203}$Pb&  5.42 $\pm$   0.02  & 4.45  $\pm$  0.05    \\
$^{204}$Pb&  1.95 $\pm$   0.18  & 1.418  $\pm$  0.14    \\
\hline
$^{202}$Bi&  0.412 $\pm$  0.237  & 0.338  $\pm$ 0.020    \\
$^{203}$Bi&  0.577 $\pm$  0.013  & 0.413 $\pm$   0.041   \\
$^{204}$Bi&  0.786 $\pm$  0.050  &  0.557  $\pm$  0.056    \\
$^{205}$Bi&  1.87 $\pm$  0.21  &  1.187$\pm$   0.133   \\
$^{206}$Bi&  1.53 $\pm$  0.012  & 1.436 $\pm$  0.144    \\
\hline
$^{206}$Po&  0.786 $\pm$  0.078  & 0.594  $\pm$  0.059    \\
\hline
$^{208}$At&  0.297 $\pm$  0.128  & 0.234 $\pm$  0.101    \\
$^{209}$At&  0.288 $\pm$  0.005  & 0.205 $\pm$   0.021   \\
$^{210}$At&  0.396 $\pm$  0.017  & 0.259 $\pm$  0.026    \\
\hline
$^{211}$Rn&  0.560 $\pm$  0.090  & 0.432 $\pm$  0.069    \\
\end{longtable}

\begin{longtable}{|c|c|c|}
\caption{Fragment cumulative and independent yields for  the reaction of  $^{204}$Hg + $^{208}$Pb at E$_{lab}$ = 1143 MeV.} \\
\hline
\textbf{Isotope}& \textbf{$\sigma$$_{CY}$ (mb)}& \textbf{$\sigma$$_{IY}$ (mb)} \\
\hline 
\endfirsthead 
\multicolumn{3}{c}{\tablename\ \thetable\ -- \textit{Continued from previous page}} \\
\hline
\textbf{Isotope}& \textbf{$\sigma$$_{CY}$ (mb)}& \textbf{$\sigma$$_{IY}$ (mb)} \\
\hline
\endhead
\hline \multicolumn{3}{c}{\textit{Continued on next page}} \\
\endfoot
\hline
\endlastfoot
$^{69}$Zn&0.264 $\pm$0.017 &  0.264$\pm$ 0.026 \\
$^{72}$Zn & 0.177 $\pm$ 0.033 & 0.176  $\pm$ 0.033 \\
\hline
$^{82}$Br & 0.329 $\pm$ 0.115 & 0.328 $\pm$ 0.115 \\
\hline
$^{86}$Y & 0.083 $\pm$ 0.012 & 0.074 $\pm$ 0.010 \\
$^{87}$Y & 0.373 $\pm$ 0.017 &0.206 $\pm$  0.020 \\
\hline
$^{91}$Sr & 0.886 $\pm$ 0.004 & 0.772  $\pm$ 0.077 \\
\hline
$^{96}$Nb & 0.325 $\pm$ 0.010 &0.325  $\pm$ 0.033 \\
\hline
$^{95}$Tc & 0.047 $\pm$ 0.013 & 0.046 $\pm$ 0.013 \\
$^{96}$Tc & 0.127 $\pm$ 0.014 & 0.127 $\pm$ 0.014 \\
\hline
$^{97}$Zr & 1.246 $\pm$ 0.016 & 1.245 $\pm$ 0.125 \\
\hline
$^{97}$Ru & 0.614 $\pm$ 0.008 & 0.565 $\pm$ 0.057  \\
\hline
$^{99}$Mo & 1.974 $\pm$ 0.016 & 1.974 $\pm$ 0.197 \\
\hline
$^{101}$Rh & 0.050 $\pm$ 0.006 & 0.025 $\pm$ 0.003 \\
\hline
$^{103}$Ru & 3.216 $\pm$ 0.006 & 0.639 $\pm$ 0.064 \\
$^{105}$Ru & 0.773 $\pm$ 0.004 &  0.659  $\pm$ 0.141 \\
\hline 
$^{111}$In & 0.053 $\pm$ 0.003 & 0.049 $\pm$ 0.005 \\
\hline 
$^{115}$Cd & 0.130 $\pm$ 0.040 & 0.085 $\pm$ 0.009 \\
\hline
$^{117}$Sb& 0.492 $\pm$ 0.029 & 0.532$\pm$ 0.053  \\
$^{120}$Sb$^{m}$& 0.101$\pm$ 0.015 &0.100  $\pm$ 0.015 \\
$^{122}$Sb& 0.201$\pm$ 0.020 & 0.200 $\pm$ 0.020 \\
$^{128}$Sb& 0.321$\pm$  0.007 & 0.088 $\pm$  0.043 \\
\hline
$^{132}$Cs&  0.214 $\pm$  0.009 & 0.214  $\pm$ 0.021    \\
\hline
$^{135}$Ba$^{m}$ & 0.759 $\pm$ 0.025 & 0.759 $\pm$ 0.076 \\
$^{140}$Ba & 0.195 $\pm$ 0.020 & 0.172 $\pm$ 0.018 \\
\hline
$^{143}$Ce&  0.210 $\pm$  0.008  &  0.177 $\pm$  0.018    \\
\hline
$^{147}$Gd&  0.118 $\pm$  0.047  &  0.114 $\pm$  0.045  \\
\hline
$^{172}$Er & 0.159 $\pm$ 0.042 & 0.159 $\pm$ 0.042 \\
\hline
$^{165}$Tm& 0.165$\pm$ 0.010 & 0.138  $\pm$ 0.014  \\
\hline
$^{169}$Lu & 0.377 $\pm$ 0.016 & 0.377 $\pm$  0.038 \\
$^{171}$Lu & 0.946 $\pm$ 0.178 & 0.752 $\pm$ 0.142 \\
\hline
$^{170}$Hf& 0.102 $\pm$  0.041 & 0.096 $\pm$ 0.039    \\
$^{173}$Hf& 0.436$\pm$ 0.045 & 0.365 $\pm$ 0.039  \\
$^{180}$Hf$^{m}$& 0.088 $\pm$  0.014 & 0.088 $\pm$  0.014 \\
\hline
$^{176}$Ta& 0.533$\pm$ 0.094 & 0.433 $\pm$ 0.076  \\
$^{177}$Ta& 0.824$\pm$ 0.039 & 0.619$\pm$ 0.062  \\
$^{178}$Ta& 0.179$\pm$ 0.050 &  0.179 $\pm$ 0.050 \\
$^{183}$Ta & 0.221 $\pm$ 0.118 &  0.192 $\pm$ 0.102 \\
\hline
$^{181}$Re& 0.555$\pm$ 0.051 & 0.454 $\pm$ 0.049  \\
$^{182}$Re& 0.686$\pm$ 0.167 & 0.253 $\pm$ 0.147  \\
$^{188}$Re& 0.716$\pm$ 0.045 & 0.705 $\pm$ 0.071  \\
$^{189}$Re & 1.19 $\pm$ 0.11 & 1.06  $\pm$ 0.11 \\
\hline
$^{183}$Os& 0.196 $\pm$ 0.036 & 0.203 $\pm$  0.021 \\
$^{185}$Os& 1.370$\pm$ 0.096 & 1.031 $\pm$ 0.103   \\
\hline
$^{186}$Ir& 0.655 $\pm$  0.147 & 0.552 $\pm$   0.124  \\
$^{188}$Ir& 0.587 $\pm$  0.083 & 0.571 $\pm$  0.081   \\
$^{190}$Ir&  0.170 $\pm$  0.052   & 0.161  $\pm$  0.016   \\
\hline
$^{188}$Pt& 0.748 $\pm$  0.100 & 0.651 $\pm$ 0.087    \\
$^{191}$Pt& 1.76 $\pm$  0.60 & 1.22 $\pm$ 0.41   \\
$^{197}$Pt$^{m}$& 0.89 $\pm$  0.17 & 0.89 $\pm$ 0.17    \\
\hline
$^{193}$Au&  0.43 $\pm$  0.07  &  0.36 $\pm$   0.06   \\
$^{194}$Au&  0.430 $\pm$  0.001  & 0.426 $\pm$  0.043    \\
$^{196}$Au&  1.652    $\pm$ 0.028  & 0.804 $\pm$   0.080    \\
$^{198}$Au&  2.414 $\pm$  0.010  & 1.932  $\pm$ 0.193   \\
$^{199}$Au&  3.25 $\pm$   0.20 & 2.65  $\pm$   0.26    \\
$^{200}$Au$^{m}$ & 0.598 $\pm$ 0.062& 0.598 $\pm$ 0.060 \\
\hline
$^{192}$Hg$^{m}$& 0.165 $\pm$  0.021 & 0.150 $\pm$  0.019  \\
$^{193}$Hg$^{m}$& 0.574 $\pm$  0.011 & 0.281 $\pm$  0.028  \\
$^{195}$Hg$^{m}$& 1.08 $\pm$  0.05 & 0.841 $\pm$ 0.085   \\
$^{197}$Hg& 1.75 $\pm$  0.13  & 1.75 $\pm$  0.17    \\
$^{203}$Hg & 19.0 $\pm$ 0.5 & 17.5 $\pm$ 1.7 \\
\hline
$^{197}$Tl&  1.53 $\pm$  0.19 & 1.239  $\pm$  0.152   \\
$^{199}$Tl&  1.51 $\pm$  0.24 &  1.05 $\pm$  0.17   \\
$^{200}$Tl&  2.92 $\pm$  0.15 &  2.09 $\pm$  0.21   \\
$^{201}$Tl&  4.70 $\pm$   0.07  & 2.83 $\pm$  0.28    \\
$^{202}$Tl& 2.73  $\pm$  0.03  & 2.72   $\pm$  0.27    \\
\hline
$^{200}$Pb& 0.39 $\pm$ 0.01 & 0.31  $\pm$  0.03   \\
$^{201}$Pb&  0.962 $\pm$  0.077  &  0.712 $\pm$  0.071   \\
$^{202}$Pb$^{m}$& 0.926 $\pm$  0.015  & 0.926  $\pm$  0.092    \\
$^{203}$Pb&  1.92 $\pm$   0.31  &  1.24 $\pm$   0.20   \\
\hline
$^{203}$Bi&  1.19 $\pm$  0.12  & 0.91 $\pm$   0.09  \\
$^{204}$Bi&  1.27 $\pm$  0.19  &  0.90  $\pm$   0.13   \\
$^{205}$Bi&  2.79 $\pm$  0.25  & 1.77 $\pm$   0.18   \\
$^{206}$Bi&  1.81 $\pm$  0.14  & 1.70 $\pm$   0.17   \\
\hline
$^{206}$Po&  1.00 $\pm$  0.15  & 0.75  $\pm$  0.11    \\
$^{207}$Po&  1.00 $\pm$  0.057 & 0.999 $\pm$  0.100    \\
\hline
$^{209}$At&  1.12 $\pm$  0.041  & 0.80 $\pm$  0.08   \\
$^{210}$At&  1.76 $\pm$  0.27  & 1.15 $\pm$ 0.18     \\
\hline
$^{211}$Rn&  0.36 $\pm$  0.09  & 0.28 $\pm$  0.07    \\
\end{longtable}

\end{document}